\documentclass[superscriptaddress,aps,epsfig,nofootinbib]{revtex4}
\usepackage{graphicx}
\usepackage{epstopdf}
\usepackage{latexsym}
\usepackage{amssymb}
\usepackage{textcomp}
\usepackage{commath}
\usepackage{amsmath}
\usepackage[center]{subfigure}
\usepackage{feynmp}
\makeatletter
\DeclareRobustCommand{\cev}[1]{%
  \mathpalette\do@cev{#1}%
}
\newcommand{\do@cev}[2]{%
  \fix@cev{#1}{+}%
  \reflectbox{$\m@th#1\vec{\reflectbox{$\fix@cev{#1}{-}\m@th#1#2\fix@cev{#1}{+}$}}$}%
  \fix@cev{#1}{-}%
}
\newcommand{\fix@cev}[2]{%
  \ifx#1\displaystyle
    \mkern#23mu
  \else
    \ifx#1\textstyle
      \mkern#23mu
    \else
      \ifx#1\scriptstyle
        \mkern#22mu
      \else
        \mkern#22mu
      \fi
    \fi
  \fi
}

\makeatother

\begin{document}
\renewcommand\thesection{\arabic{section}}
\renewcommand\thesubsection{\thesection.\arabic{subsection}}
\renewcommand\thesubsubsection{\thesubsection.\arabic{subsubsection}}
 \newcommand{\bq}{\begin{equation}}
 \newcommand{\eq}{\end{equation}}
 \newcommand{\bqn}{\begin{eqnarray}}
 \newcommand{\eqn}{\end{eqnarray}}
 \newcommand{\nb}{\nonumber}
 \newcommand{\lb}{\label}
 \title{Lorentz Gauge Theory of Gravity in Electron-–Positron Colliders}
 \author{Ahmad Borzou}
\email{ahmad_borzou@baylor.edu}
\affiliation{EUCOS-CASPER, Physics Department, Baylor University, Waco, TX 76798-7316, USA}
\affiliation{Department of Physics, Isfahan University of Technology, Isfahan, 84156-83111, Iran}
\author{Gerald Cleaver}
\email{gerald_cleaver@baylor.edu}
\affiliation{EUCOS-CASPER, Physics Department, Baylor University, Waco, TX 76798-7316, USA}
\author{Behrouz Mirza}
\email{b.mirza@cc.iut.ac.ir}
\affiliation{Department of Physics, Isfahan University of Technology, Isfahan, 84156-83111, Iran}

\date{\today}

\begin{abstract}
Lorentz gauge theory (LGT) is a feasible candidate for theory of quantum gravity in which routine field theory calculations can be carried out perturbatively without encountering too many divergences. In LGT spin of matter also gravitates. The spin-generated gravity is expected to be extremely stronger than that generated by mass and could be explored in current colliders. In this article the observable signals of the theory in an electron-positron collider is investigated. We specifically study pair annihilation into two gravitons, and LGT corrections to processes like $e^-+e^+\rightarrow \mu^-+\mu^+$ and $e^-+e^+\rightarrow e^-+e^+$.
\end{abstract}

\maketitle

\section{Introduction}
Gravity is expected to be quantized at small scales \cite{Albers2008, Kiefer2007}. However, there is yet no theory of quantum gravity that meets all the expectations. Several directions have been tried so far. Canonical quantization of general relativity \cite{Dewitt,Arnowitt2008} is perhapse the oldest approach. String theory \cite{Polchinski} and loop quantum gravity \cite{Ashtekar,Rovelli} are other well familiar endeavors.
LGT \cite{Borzou2016} should be added to the list of candidates for a quantum theory of gravity.  
It is a Yang-Mills theory based on internal Lorentz symmetry of fermions in which the metric is not dynamic and the energy-momentum tensor is not the source of gravity. In LGT interactions of gravitons with fermionic spins are expected to become significant at energies much lower than the Planck scale.
On the other hand, several electron-positron accelerators like the large electron positron collider (LEP) \cite{LEP}, and the SLAC linear collider (SLC) \cite{SLC}, have already collected lots of data. Moreover, there are ongoing studies for high luminosity colliders with the center of mass energies in the TeV range. Examples are the international linear collider (ILC) \cite{ILC} and the compact linear collider (CLIC) \cite{CLIC}. 
Therefore, searches for LGT signals in the current or near future experiments are well motivated and in the present paper we are going to study these interactions and their signatures in electron-positron colliders.   

In this paper, the following conventions are adopted. For the sake of simplicity the study is restricted to the important case of relativistic collisions where masses can be neglected altogether. Moreover, calculations are all restricted to the center of mass frame in which all the incoming and outgoing particles have the same energy. 
To reduce the effects of the weak interactions the center of mass energy is restricted to be below the mass of the Z boson. We further assume that the incoming electrons and positrons are moving in the positive and negative z directions respectively. Due to the cylindrical symmetry in such interactions, calculations are restricted in the x-z plane. Therefore, the scatterings can be characterized with the outgoing fermion's angle $\theta$. 

In this article we first summarize Lorentz gauge theory in section \ref{LGTsec}. Next, in sections \ref{eemumuSec} and \ref{bhabhaSec} quantum corrections from LGT into $e^-+e^+\rightarrow \mu^-+\mu^+$ and $e^-+e^+\rightarrow e^-+e^+$ are investigated respectively. In section \ref{gravitonCreationSec}, after studying the gravitational plane waves and deriving their physical modes, the process of pair annihilation into gravitons is investigated. A conclusion is given in the end in section \ref{conclusionSec}.

\section{Lorentz gauge theory of gravity}
\lb{LGTsec}
General relativity was first proposed to remain invariant only under general coordinate transformations
\bqn
\tilde{x}^{\mu} = x^{\mu} + \xi^{\mu}(x),
\eqn
where $\xi^{\mu}(x)$ is an arbitrary vector and can be written as
\bqn
\xi^{\mu}(x) = \Lambda^{\mu}_{~\nu}(x)x^{\nu} + \left(\xi^{\mu}(x) - \Lambda^{\mu}_{~\nu}(x)x^{\nu}\right).
\eqn
Here $\Lambda^{\mu}_{~\nu}(x)$ indicates a Lorentz transformation and the last term is a translation, i.e., the whole change is a Poincare transformation. 
This original version of general relativity is good as far as fermions can be neglected. The reason is that the group of general linear $4\times 4$ matrices has no representation that behaves like spinors under the Lorentz group. 
In order to incorporate fermionic fields in general theory of relativity, people have utilized the concept of Minkowskian tangent spaces defined at each point of a semi-Riemannian space-time. 
This opens the possibility of defining the spinor fields in these tangent spaces and requiring that their Lagrangian remains invariant under Lorentz transformations that are solely defined in the new spaces. Matter Lagrangians have to remain invariant under any general coordinate transformation in the space-time and also under any Lorentz transformation in the tangent spcaes.  
Hence, GR can be regarded as a theory that is invariant under \cite{WeinbergGRBook}\footnote{look for the paragraph: ``There are now two invariance principles which must be met in constructing a suitable matter action $I_{\text{M}}$: (A) The action must be generally covariant, $\cdots$. (B) $\cdots$ with respect to Lorentz transformation $\Lambda^{\alpha}_{~\beta}(x)$ that can depend on position in space-time $\cdots$. These two invariance principles lead to a dual classification of physical quantities. A coordinate scalar or coordinate tensor transforms as a scalar or a tensor under changes in the coordinate system. A Lorentz scalar or Lorentz tensor or Lorentz spinor transforms according to a rule like $\cdots$ under changes in the choice of the locally inertial coordinate frame''.  }
\bqn
\text{General Covariance}\otimes\text{Lorentz}.\nb
\eqn
Consequently, physical objects belong to one of the two groups. The metric and the vector bosons belong to space-time and are coordinate tensors while fermions belong to the Lorentz space and objects like ``$\bar{\psi}(\text{combination of }\gamma^i)\psi$'' are Lorentz tensors. 
The Lorentz spaces are connected with space-time through a set of four vector fields called the tetrad. The tetrad can be decomposed to its space-time and Lorentz components at each point of space-time 
\bq
e_{i \mu}=\eta^{\bar{k} \bar{l}} e_{i \bar{k}}e_{\bar{l}\mu},
\eq
where components of the objects in the free falling frame are referred to with a bar and  
\bqn
e_{\bar{l}\mu} &\equiv& \hat{e}_{\bar{l}}\cdot \hat{e}_{\mu}\nb\\
e_{i\bar{k}} &\equiv& \hat{e}_i \cdot \hat{e}_{\bar{k}},
\eqn
while $\hat{e}_i$ are the unit vectors in the Lorentz space at that point, $\hat{e}_{\bar{l}}$ are the unit vectors carried by the free falling observer at that point, and $\hat{e}_{\mu}$ are the unit vectors tangent to coordinates at that point. The general covariance leads to conservation of energy-momentum and also angular-momentum. The Lorentz invariance on the other hand leads to a conserved current with angular momentum dimension to which Lagrangians of Bosons or vacuum energy  do not contribute. The current is derived in appendix A. 
Having two independent sets of conserved currents, we have to decide which one is the source of gravity. In GR the energy-momentum tensor is chosen as the source of gravity. That corresponds to assuming that the space-time component of the tetrad is dynamic, i.e. $\delta e_{i \mu}=\eta^{\bar{k} \bar{l}} e_{i \bar{k}}\delta e_{\bar{l}\mu}$.
This however leads to several unsolved problems. Examples are the cosmological constant problem \cite{Weinberg1989,Martin2012}, the problem of time \cite{Kiefer2013,Isham1993}, and nonrenormalizability \cite{tHooft1974,Deser1999}. 
In Lorentz gauge theory of gravity (LGT), it is assumed that the conserved current of the Lorentz space is the source of gravity which means the Lorentz component of the tetrad is dynamic $\delta e_{i \mu}=\eta^{\bar{k} \bar{l}} \delta e_{i \bar{k}} e_{\bar{l}\mu}$. This enables us to define a Yang-Mills theory whose equations determine the spin connections and the latter is sufficient to determine the tetrad and therefore the metric and the metric-compatible Christoffel symbols. A simple dimensional analysis shows that the coupling constant of LGT is dimensionless
compared to that of GR which has negative two dimension. Therefore, LGT is expected to have a much better high energy behavior than GR. 
LGT is formally defined with the following action \cite{Borzou2016}
\bqn
\lb{action}
S&=&\int e d^4x\Big[{\cal{L}}_{A}+{\cal{L}}_{M}+{\cal{L}}_{C}\Big],
\eqn
where ${\cal{L}}_{A}$ is the gauge field's Lagrangian and is given by 
\bqn
\lb{NLA}
{\cal{L}}_{A}&=&-\frac{1}{4g^2}F_{\mu \nu ij}F^{\mu\nu ij},
\eqn
where the strength tensor is defined in just the same way as in any Yang-Mills theory
\bq
\lb{strength}
F_{\mu\nu ij}=g\partial_{\nu}A_{ij\mu}-g\partial_{\mu}A_{ij\nu}+g^2A_{i~~\mu}^{~m}A_{mj\nu}-g^2A_{i~~\nu}^{~m}A_{mj\mu}.
\eq 
It should be mentioned that the symmetries of the theory allow other terms in the Lagrangian as well, which can be found in \cite{Hayashi01091980, PhysRevD.80.104031}. But, they lead to interactions that are not familiar from the standard model and our preliminary evaluation suggests that they are not renormalizable and are abandoned for this reason.  ${\cal{L}}_{M}$ is the Lagrangian of the standard model while ${\cal{L}}_{C}$ is just the tetrad postulate times a Lagrange multiplier
\bqn
\lb{LagrangeMulti}
{\cal{L}}_{C}&=&S^{\mu \nu i}D_{\mu}e_{i \nu}\nb\\
&=&0.
\eqn
The reason for the latter Lagrangian is that the tetrad in the action should be expressed in terms of the spin connections, which itself requires solving an integral equation. To avoid this cumbersome task, the dependence of the tetrad on the spin connections is entered into the action as a constraint and the two fields are treated independently afterward. The field equations are found by varying the action in the Lorentz spaces, i.e. under which coordinate tensors like the metric remain unchanged. A variation with respect to the tetrad gives a constraint equation
\bqn
\lb{fieldconstraint}
D_{\mu}S^{\mu \nu i}=\frac{\delta{\cal{L}}_M}{\delta e_{i \nu}},
\eqn
where $\frac{\delta{\cal{L}}_M}{\delta e_{i \nu}}$ is just the energy-momentum tensor. A variation with respect to the spin connections on the other hand gives the dynamical field equations
\bqn
\lb{fielddynamic}
D_{\nu}F^{\mu \nu ij}=-\frac{\delta{\cal{L}}_{M}}{\delta A_{ij \mu}}+S^{\mu \nu [i}e^{j]}_{~\nu},
\eqn
where the first term on the right hand side is just the spin angular momentum of fermions. Therefore, it is the second term that gives rise to the Newtonian gravity. 

Comparing the two source terms, it is seen that the coupling constant of LGT appears alone in the first term but is multiplied by a vector, with dimension of length, in the second one.
To arrive at the Newtonian gravity in the classical regimes, we have to assume that the latter length belongs to the Planck scale. 
By absorbing the small length, labeled $\delta$, into the coupling constant of LGT, a classically effective version is developed  \cite{Borzou2016_2} in which the effective coupling constant is Newton's gravitational constant
\bqn
\lb{NewtonConst}
G=\frac{g \delta ^2}{40 \pi}.
\eqn
Since this equation is derived after assuming that $\delta \sim \sqrt{G}$, we can conclude that the strength of $g$ is comparable to the coupling constants in the standard model, i.e. $g \sim 1$. 
On the other hand, the coupling constant of the first source term in equation \eqref{fielddynamic} is $g$, and is not multiplied by $\delta$. It suggests that interactions contained in this term become significant at energies much lower than the Planck mass. Study of these interactions is the subject of the present paper where the second term on the right hand side of equation \eqref{fielddynamic} is totally neglected. 
After dropping the irrelevant interactions and fixing the gauge by $\partial^{\nu}A_{ij \nu}=0$, Feynman rules for LGT in the momentum space is as follows \cite{Borzou2016}. The propagator is just the inverse of $\partial ^2$ in the linear field equations, given in equation \eqref{linearFieldEq},
\begin{fmffile}{propag}
\bqn  
  \parbox{27mm}{\begin{fmfgraph*}(50,4)
    \fmfleft{l}
    \fmf{wiggly,label=$q$}{l,r}
    \fmfright{r}
    \fmflabel{$ij\mu$}{l}
    \fmflabel{$mn\nu$}{r}
     \end{fmfgraph*}}
& = & -\frac{i}{2}\frac{\eta_{\mu\nu}\Big(\eta_{mi}\eta_{nj}-\eta_{mj}\eta_{ni}\Big)}{q^2+i \varepsilon}.
\eqn
\end{fmffile}
The only important self interaction at the tree level, which is the subject of the present study, is obtained by dropping $g^2$ terms and varying the Lagrangian three times with respect to $A_{ij\mu}$

\bqn
    &&~~~~~~~~~~~~~~~~~~~~~~\begin{fmffile}{A3Vertex}
      \setlength{\unitlength}{1.cm}
      \parbox{30mm}{\begin{fmfgraph*}(3.5,3.5)
  \fmfright{i1,i2}
  \fmfleft{o}
  \fmf{wiggly,label=$k_2$,label.side=right}{i1,v}
  \fmf{wiggly,label=$k_1$}{o,v}
  \fmf{wiggly,label=$k_3$}{i2,v}
        \fmflabel{$i_2j_2\mu_2$}{i1}
        \fmflabel{$i_3j_3\mu_3$}{i2}
        \fmflabel{$i_1j_1\mu_1$}{o}
      \end{fmfgraph*}}
    \end{fmffile}\nb\\
    ~\nb\\
    =&&X^{i_1j_1\mu_1,i_2j_2\mu_2,i_3j_3\mu_3}(k_2,k_3)\nb\\
    =&&-\frac{g}{2}\left(\left(\left(\eta^{i_1i_3}\eta^{j_1j_2}\eta^{i_2j_3}-i_1\leftrightarrow j_1\right)-i_2\leftrightarrow j_2\right)-i_3\leftrightarrow j_3\right)\times\nb\\
    &&~~~~~\Bigg(2k_3^{\mu_2}\eta^{\mu_3\mu_1}-k_3^{\mu_1}\eta^{\mu_2\mu_3}-2k_2^{\mu_3}\eta^{\mu_2\mu_1}+k_2^{\mu_1}\eta^{\mu_2\mu_3}\Bigg).
\eqn
Also, the only relevant matter interaction is obtained after varying the matter Lagrangian, given in equation \eqref{MatterLagrange}, with respect to $\bar{\psi}$, $\psi$, and $A_{ij\mu}$ and putting $\delta=0$, equivalent to assuming $\frac{\delta e_{i\mu}}{\delta A_{mn\nu}}=0$,
\bqn
 \lb{vert_low}
\begin{fmffile}{spin}
 \parbox{31mm}{\begin{fmfgraph*}(60,60)
  \fmfleft{i1,i2}
  \fmfright{o}
  \fmf{fermion}{i1,v}
  \fmf{wiggly}{v,o}
  \fmf{fermion}{i2,v}
  \fmflabel{$mn\nu$}{o}
 \end{fmfgraph*}}
   = g\delta_k^{\nu}\{\gamma^k,S^{mn}\},
\end{fmffile}
\eqn
where the tetrad in flat background is shown with a delta and $\{a,b\}\equiv \frac{1}{2}(a\cdot b +b\cdot a)$.

So far we have found two exact vacuum solutions for LGT, namely the Schwarzschild and the de Sitter spaces. The former is crucial to passing the solar system tests of gravity while the latter is needed to explain inflationary expansions in the very early times and in the late times in the universe. It should be noted that the solution holds for when there is no matter, nor dark energy, in the universe. Therefore, LGT does not need dark energy to explain such expansions. Moreover, in LGT transition from decelerating to accelerating expansion in the late times is spontaneous and is driven by geometrical terms. Also, unlike in GR, in LGT the vacuum energy does not gravitate \cite{Borzou2017}. See appendix A for more details. Therefore, there will be no more contradiction with quantum field theory that predicts a very large value for the vacuum energy. All the supplementary materials that are needed to reproduce our results in this and preceding papers are gathered in a repository that can be found at \cite{LGTRepo}.

\section{LGT corrections to $e^-+e^+\rightarrow \mu^-+\mu^+$}
\lb{eemumuSec}
The effects of LGT in processes like $e^-+e^+\rightarrow \mu^-+\mu^+$ are investigated in this section. Because of the similarity between the LGT diagram in equation \eqref{vert_low} and the only interaction in QED, it is already clear that there are corrections from LGT into such processes. 
The Feynman diagrams for this interaction from both QED and LGT are given in the following
\bqn
    &&\begin{fmffile}{LeptonPairProduction}
      \setlength{\unitlength}{1.cm}
      \parbox{30mm}{\begin{fmfgraph*}(5,3)
        \fmfleft{ld,lu}
        \fmfright{rd,ru}
        \fmf{fermion,label=$e^-$}{ld,w1}
        \fmf{fermion,label=$l^-$}{w0,lu}
        \fmf{boson,label=$A,,\gamma$}{w1,w0}
        \fmf{fermion,label=$e^+$}{w1,rd}
        \fmf{fermion,label=$l^+$}{ru,w0}
        \fmflabel{$s'$}{lu}
        \fmflabel{$r'$}{ru}
        \fmflabel{$r$}{rd}
        \fmflabel{$s$}{ld}
      \end{fmfgraph*}}
    \end{fmffile}\nb\\
\eqn
where the internal line is a photon $\gamma$ in QED and a graviton $A_{ij\mu}$ in LGT. The amplitude corresponding with the two diagrams is given by
\bqn
i{\cal{M}}^{s r s'r'}&=&
-e^2\bar{v}^{r}(k)\delta^{k\mu}\gamma_ku^s(p)\frac{-i\eta_{\mu\nu}}{(p+p')^2}\bar{u}^{s'}(p')\delta^{l\nu}\gamma_lv^{r'}(k')\nb\\
&+&g^2\bar{v}^{r}(k)\delta^{k\mu}\{\gamma_k,S^{ij}\}u^s(p)\frac{-\frac{i}{2}\eta_{\mu\nu}\Big(\eta_{mi}\eta_{nj}-\eta_{mj}\eta_{ni}\Big)}{(p+p')^2}\bar{u}^{s'}(p')\delta^{l\nu}\{\gamma_l,S^{mn}\}v^{r'}(k').
\eqn
After a long but straightforward calculation the non-zero components of the amplitude turn out to be
\begin{align}
{\cal{M}}^{1212}&=\left(\frac{3}{2}g^2-2e^2\right)\cos^2(\frac{\theta}{2}),&  
{\cal{M}}^{2112}&=\left(\frac{3}{2}g^2+2e^2\right)\sin^2(\frac{\theta}{2}),\nb\\
{\cal{M}}^{1221}&=\left(\frac{3}{2}g^2+2e^2\right)\sin^2(\frac{\theta}{2}),& 
{\cal{M}}^{2121}&=\left(\frac{3}{2}g^2-2e^2\right)\cos^2(\frac{\theta}{2}).
\end{align}
Since particle detectors are usually blind to polarization, we are also interested in unpolarized amplitude. Hence, a sum on the outgoing spins and an average over the incoming ones is in order
\begin{figure}[h]
\centering
\includegraphics[width=14cm]{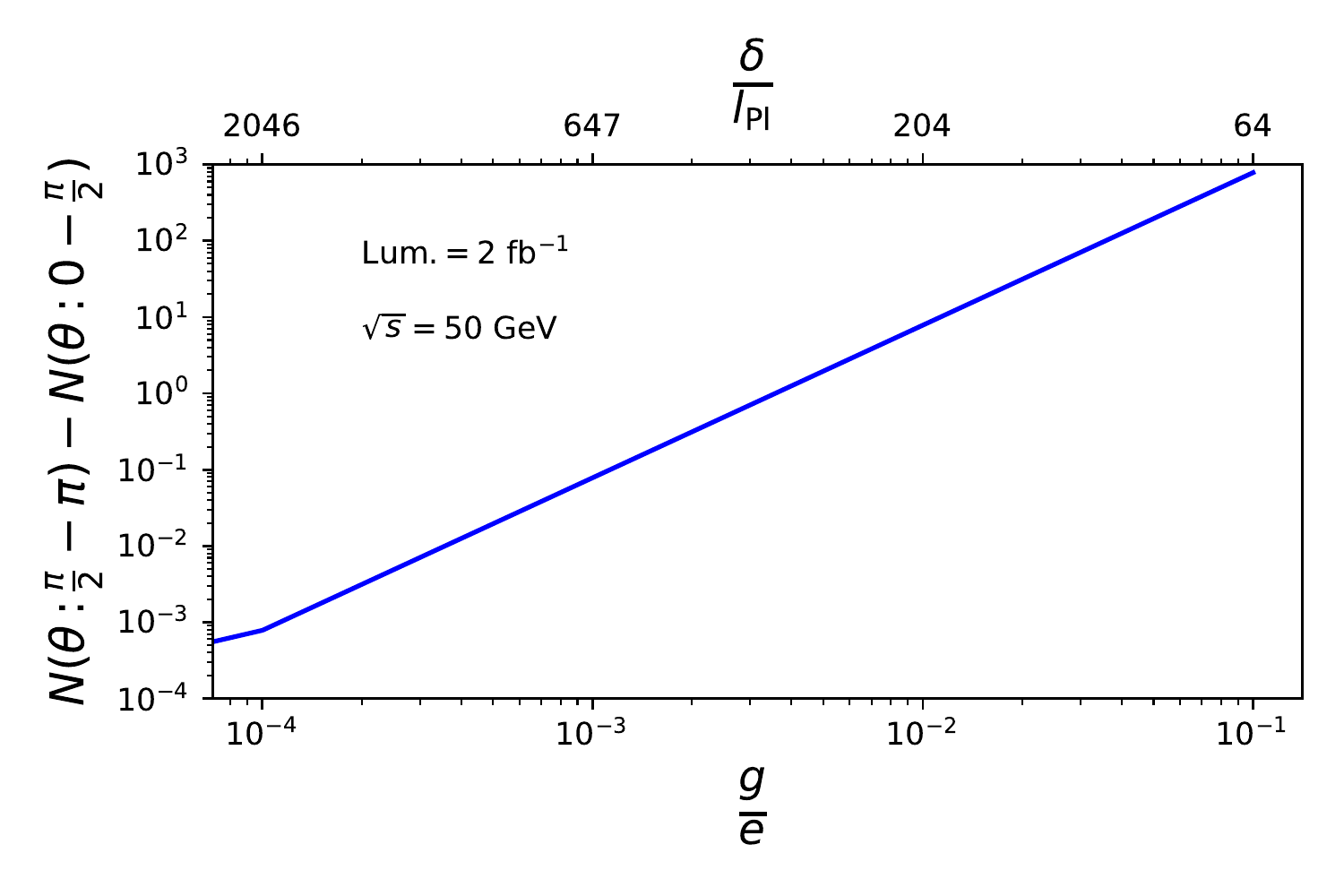}
\caption{There exists an asymmetry in the observed number of events in the left and right sides of the detector. The asymmetry is given as a function of the ratio of LGT to QED coupling constants. The top x-axis shows the ratio of the scale $\delta$ beyond which other gravity-matter interactions become significant, over the Planck length.}
\lb{ANevents}
\vspace{0.75cm}
\end{figure}
\bqn
\frac{1}{4}\sum_{\text{spin}}|{\cal{M}}|^2=\left(e^4+\frac{9}{16}g^4\right)\left(1+\cos^2(\theta)\right)-3e^2g^2\cos(\theta).
\eqn

Therefore the differential cross section in the center of mass frame takes the following form
\bqn
\lb{eemumuCross}
\frac{d\sigma}{d\Omega}=\frac{\alpha^2}{4E^2_{\text{cm}}}\left(\left(1+\frac{9}{16}r^4\right)\left(1+\cos^2(\theta)\right)-3r^2\cos(\theta)\right),
\eqn
where $\alpha=\frac{e^2}{4\pi}$ is the QED coupling constant, while $r\equiv \frac{g}{e}$ and is zero if LGT is switched off.
While the QED term of the cross section is symmetric between the forward and backward hemispheres of particle detectors, separated by the $\theta=\frac{\pi}{2}$ plane, the LGT term is not. Fermions that undergo gravitational interactions tend to scatter more into the backward region. Therefore, one viable search for LGT is to subtract the number of events in the backward hemisphere from those in the forward one and see if the number is different from zero. 
The difference in the total number of observed events is given by 
\bqn
N(\theta:\frac{\pi}{2}-\pi)-N(\theta:0-\frac{\pi}{2})=\text{luminosity}\times2\pi\left(\int_{\frac{\pi}{2}}^{\pi}\sin(\theta)\frac{d\sigma}{d\Omega}d\theta-\int_0^{\frac{\pi}{2}}\sin(\theta)\frac{d\sigma}{d\Omega}d\theta\right),
\eqn 
which is plotted in figure \ref{ANevents} as a function of the coupling constant of LGT over that of QED at a center of mass energy of 50 GeV and luminosity of 2 fb$^{-1}$.
If the value of LGT coupling constant is measured one day, we can use equation \eqref{NewtonConst} to find the value of the scale $\delta$, beyond which the neglected interactions are also important. The top x-axis in the same figure shows this scale divided by the Planck length. 
Finally we use a simple $\chi^2(r)$ fit to set an upper limit on $r=\frac{g}{e}$ 
\bqn
\lb{chi2}
\chi^2(r)=\text{luminosity}\times\frac{\sigma_{\text{LGT}}^2}{\sigma_{\text{QED}}},
\eqn
where $\sigma_{\text{QED}}$ refers to the first term in equation \eqref{eemumuCross} and $\sigma_{\text{LGT}}$ refers to the second one, both integrated over the whole phase space.
A $95\%$ CL upper limit on $r$ will be obtained by requiring that $\chi^2(r)<0.004$. Solving the inequality for a center of mass energy of 50 GeV and luminosity of 2 fb$^{-1}$
\bqn
r<0.014.
\eqn

\section{LGT corrections to $e^-+e^+\rightarrow e^-+e^+$}
\lb{bhabhaSec}
This section is devoted to Bhabha scattering to which not only the S-channel amplitude of the previous section, but also a T-channel amplitude given by the following diagram contributes

\bqn
i{\cal{M}}^{s~r~s'r'}&=& 
    \begin{fmffile}{Potential_Graph}
      \setlength{\unitlength}{1.cm}
      \parbox{30mm}{\begin{fmfgraph*}(5,3)
        \fmfleft{ld,lu}
        \fmfright{rd,ru}
        \fmf{fermion,label=$e^-$}{ld,w1}
        \fmf{fermion,label=$e^-$,label.side=right}{w1,lu}
        \fmf{boson,label=$A,,\gamma$}{w1,w0}
        \fmf{fermion,label=$e^+$,label.side=right}{w0,rd}
        \fmf{fermion,label=$e^+$,label.side=right}{ru,w0}
        \fmflabel{$s'$}{lu}
        \fmflabel{$r'$}{ru}
        \fmflabel{$r$}{rd}
        \fmflabel{$s$}{ld}
      \end{fmfgraph*}}
    \end{fmffile}
     \nb\\
     ~\nb\\
     ~\nb\\
     &=&-\left(ie\right)^2\left(\bar{u}^{s'}(p')\delta^{k\mu}\gamma_k u^s(p)\frac{-i\eta_{\mu\nu}}{(p-p')^2}\bar{v}^{r}(k)\delta^{l\nu}\gamma_l v^{r'}(k')\right)\nb\\
     &&-g^2\left(\bar{u}^{s'}(p')\delta^{k\mu}\{\gamma_k,S^{ij}\}u^s(p)\frac{-\frac{i}{2}\eta_{\mu\nu}\Big(\eta_{mi}\eta_{nj}-\eta_{mj}\eta_{ni}\Big)}{(p-p')^2}\bar{v}^{r}(k)\delta^{l\nu}\{\gamma_l,S^{mn}\}v^{r'}(k')\right).
\eqn
Therefore, the non-zero amplitudes for polarized beams read
\begin{align}
{\cal{M}}^{1111}&=\left(2e^2+3g^2\right)\csc^2\left(\frac{\theta}{2}\right),&  
{\cal{M}}^{1212}&=\left(2e^2-3g^2\right)\cot^2\left(\frac{\theta}{2}\right),\nb\\
{\cal{M}}^{2121}&=\left(2e^2-3g^2\right)\cot^2\left(\frac{\theta}{2}\right),& 
{\cal{M}}^{2222}&=\left(2e^2+3g^2\right)\csc^2\left(\frac{\theta}{2}\right).\nb\\
\end{align}
After adding these T-channel amplitudes to those from  the S-channel and averaging over initial polarizations and summing the final ones we get
\bqn
\frac{1}{4}\sum_{\text{spin}}|{\cal{M}}|^2
&=&2e^4\left(
\frac{1 + \cos^4\left(\frac{\theta}{2}\right)}{\sin^4\left(\frac{\theta}{2}\right)} + \frac{1 + \cos^2\left(\theta\right)}{2} - \frac{2 \cos^4\left(\frac{\theta}{2}\right)}{\sin^2\left(\frac{\theta}{2}\right)}
\right)\nb\\
&+&3e^2g^2\left(5+2\cos\left(\theta\right)+\cos\left(2\theta\right)\right)\csc^{2}\left(\frac{\theta}{2}\right)
+\frac{9}{64}g^4\left(7+\cos\left(2\theta\right)\right)^2\csc^{4}\left(\frac{\theta}{2}\right).
\eqn
The differential cross section of the Bhabha scattering for three different LGT couplings' strengths is shown in figure \ref{BhabhaFig} for a center of mass energy of 50 GeV. 
To set an upper limit on the coupling constant of LGT, equation \eqref{chi2} can be used again. It turns out that with the same center of mass energy and luminosity, Bhabha scattering will set a tighter limit of 
\bqn
r<0.006.
\eqn
\begin{figure}[h]
\centering
\includegraphics[width=14cm]{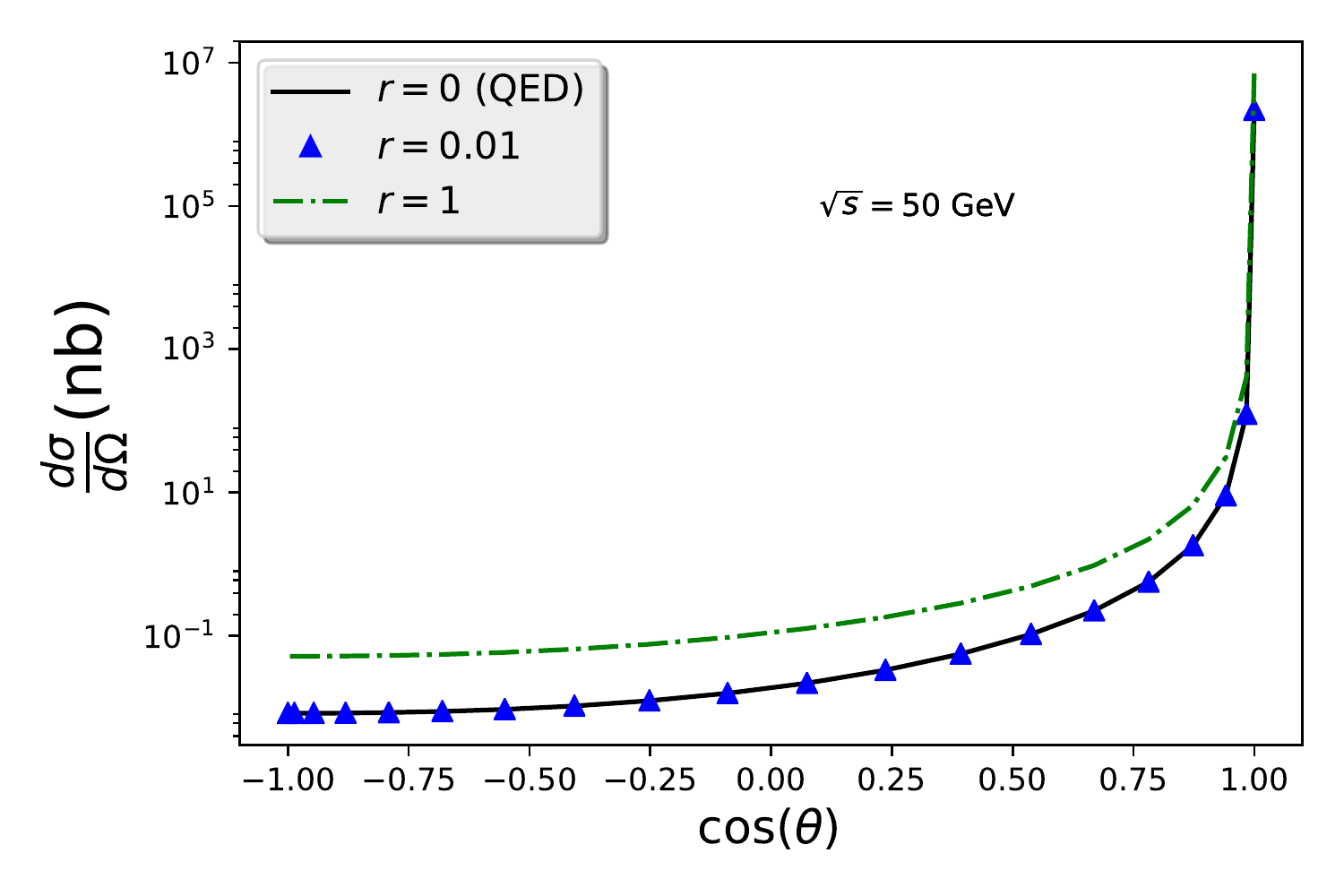}
\caption{The differential cross section of Bhabha scattering as a function of the scattering angle for three different ratios of LGT to QED strengths. Here the ratio of LGT coupling constant over that of QED is shown with $r$.}
\lb{BhabhaFig}
\vspace{0.75cm}
\end{figure}

\section{Pair Annihilation into Gravitons}
\lb{gravitonCreationSec}

So far we have studied the LGT corrections to the interactions that are dominated by QED. In this section we would like to focus on purely LGT dominated events of pair annihilation into gravitons. 
This work is not possible until the Feynman rules for gravitons is worked out. In the following, first a plane gravitational wave is studied and physical and non-physical modes are investigated. Next, the amplitudes of interest are calculated by drawing the relevant Feynman diagrams. 

\subsection{Gravitational Plane Wave and Feynman Rules for Gravitons}

A plane gravitational wave that is freely propagating in space should satisfy the linearized LGT field equations 
\bqn
\lb{linearFieldEq}
\partial^2 A_{ij\mu}-\partial^{\nu}\partial_{\mu}A_{ij\nu}=0.
\eqn
LGT is invariant under both local Lorentz transformations in the Lorentz spaces and arbitrary change of coordinates in the space-time given by 
\bqn
&&\Lambda_i^{~m}=\delta_i^{m}+\omega_i^{~m},\nb\\
&&\frac{\partial x^{\nu}}{\partial \tilde{x}^{\mu}}=\delta^{\nu}_{\mu}+\partial_{\tilde{\mu}}\xi^{\nu},
\eqn
respectively where $\omega_i^{~m}$ and $\xi^{\mu}$ are arbitrary but small parameters.
Under these two transformations the propagating field in LGT changes as 
\bqn
\lb{linearTransform}
\tilde{A}_{ij\mu}&=&\frac{\partial x^{\nu}}{\partial \tilde{x}^{\mu}}
\left(\Lambda_i^{~m}\Lambda_j^{~n}A_{mn\nu}+\Lambda_j^{~n}\partial_{\nu}\Lambda_{in} \right)\nb\\
&=&A_{ij\mu}+\partial_{\mu}\omega_{ij}+{\cal{O}}(2).
\eqn
Due to this gauge freedom, we can choose to work with the class of divergence-free spin connections. Therefore, the field equations reduce to 
\bqn
\lb{linearFieldEq}
\partial^2A_{ij\mu}=0,\nb\\
\partial^{\nu}A_{ij\nu}=0.
\eqn
These equations are satisfied if a free plane wave is described by
\bqn
\lb{planewavegauge}
&&A_{ij\mu}=e_{ij\mu}e^{ik\cdot x}+e_{ij\mu}^*e^{-ik\cdot x},\nb\\
&&e_{ij\mu}k^{\mu}=0.
\eqn
Since $A_{ij\mu}$ is antisymmetric in the two Lorentz indices, it has at most 24 independent components. However, the condition above kills 6 of them. 
To see which of the remaining components are physical, the field should be transformed according to equation \eqref{linearTransform} after utilizing 
\bqn
&&\omega_{ij}=ie_{ij}e^{ik\cdot x}-ie_{ij}^*e^{-ik\cdot x},\nb\\
&&e_{ij}=
\begin{cases}
1&\text{i}<\text{j}\\
0&\text{i}=\text{j}\\
-1&\text{i}>\text{j}
\end{cases}.
\eqn

The transformation is therefore equivalently represented by the following 
\bqn
\lb{gaugeTransOfPlaneWave}
\tilde{e}_{ij\mu}=e_{ij\mu}-k_{\mu}e_{ij}.
\eqn
Without loss of generality, the wave is assumed to propagate in the z direction, i.e. $k^{\mu}=(k,0,0,k)$. Hence, the latter equation implies
\bqn
\tilde{e}_{ij3}&=&e_{ij3}-ke_{ij},\nb\\
\tilde{e}_{ij2}&=&e_{ij2},\nb\\
\tilde{e}_{ij1}&=&e_{ij1},
\eqn
while equation \eqref{planewavegauge} leads to
\bqn
\tilde{e}_{ij0}=-\tilde{e}_{ij3}. 
\eqn
It is now easy to see that $e_{ij1}$ and $e_{ij2}$ can not be eliminated under any transformation. This means the gravitational waves are transverse just like the electromagnetic waves are. The transverse plane can be spanned with two orthogonal unit vectors $\frac{1}{\sqrt{2}}\left(0,1,\pm i,0\right)$. 
The twelve remaining components can therefore be written as 
\bqn
e_{ij\mu}^1\left(k\hat{z}\right)&=&\frac{e_{ij}}{2\sqrt{2}}\left(0,1,+i,0\right),\nb\\
e_{ij\mu}^2\left(k\hat{z}\right)&=&\frac{e_{ij}}{2\sqrt{2}}\left(0,1,-i,0\right).
\eqn

The study presented above suggests two Feynman rules for gravitons
\begin{fmffile}{gravitonIn}
\bqn  
  \parbox{20mm}{\begin{fmfgraph*}(50,4)
    \fmfleft{l}
    \fmf{wiggly,label=$\longleftarrow k$,label.side=left}{l,r}
    \fmfright{r}
    \fmfdot{l}
    \fmflabel{$r$}{r}
     \end{fmfgraph*}}
&~ = & e_{ij\mu}^r\left(\vec{k}\right),\nb\\
\parbox{20mm}{\begin{fmfgraph*}(50,4)
    \fmfleft{l}
    \fmf{wiggly,label=$k \longrightarrow$}{l,r}
    \fmfright{r}
    \fmfdot{l}
    \fmflabel{$r$}{r}
     \end{fmfgraph*}}
&~ = & e_{ij\mu}^{*r}\left(\vec{k}\right).
\eqn
\end{fmffile}
According to equation \eqref{linearTransform}, $A_{ij\mu}$ does not transform like a tensor under Lorentz transformations and as a result it is not possible to define spin angular momentum for gravitons. This may look odd at first because every other known particle carries spin angular momentum. But, the spin of those particles is to leave the quantum amplitudes invariant under global Lorentz transformations and the sole reason for the existence of the spin connections is to locally preserve the same symmetry. 
In the following when Feynman diagrams with external gravitons are considered, we conservatively will make sure that the amplitudes remain invariant under Lorentz transformations by checking that the following (which is similar to the Ward identity) holds
\bqn
{\cal{M}}={\cal{M}}^{ij\mu}e_{ij\mu}={\cal{M}}^{ij\mu}\tilde{e}_{ij\mu},
\eqn
where $\tilde{e}_{ij\mu}$ is given by equation \eqref{gaugeTransOfPlaneWave}.

\subsection{Computation of Cross Section}
Now that Feynman rules for gravitons are known, we can draw the diagrams for pair annihilation into two gravitons in all S, T, and U channels and calculate the amplitude as usual. The diagrams are shown below

\bqn
~\nb\\
    &&\begin{fmffile}{LeptonAnnihilation}
      \setlength{\unitlength}{1.cm}
      \parbox{50mm}{\begin{fmfgraph*}(5,3)
        \fmfleft{ld,lu}
        \fmfright{rd,ru}
        \fmf{fermion,label=$e^-$}{ld,w1}
        \fmf{boson,label=$A$}{w0,lu}
        \fmf{boson,label=$A$}{w1,w0}
        \fmf{fermion,label=$e^+$}{w1,rd}
        \fmf{boson,label=$A$}{ru,w0}
        \fmflabel{$s'$}{lu}
        \fmflabel{$r'$}{ru}
        \fmflabel{$r$}{rd}
        \fmflabel{$s$}{ld}
      \end{fmfgraph*}
      }
      +
      \parbox{50mm}{\begin{fmfgraph*}(5,3)
        \fmfleft{ld,lu}
        \fmfright{rd,ru}
        \fmf{fermion,label=$e^-$}{ld,wl}
        \fmf{fermion,label=$e^+$,label.side=right}{wr,rd}
        \fmf{fermion}{wl,wr}
        \fmf{boson,label=$A$,label.side=right}{wl,lu}
        \fmf{boson,label=$A$}{ru,wr}
        \fmflabel{$s'$}{lu}
        \fmflabel{$r'$}{ru}
        \fmflabel{$r$}{rd}
        \fmflabel{$s$}{ld}
      \end{fmfgraph*}
      }
      +
      \parbox{50mm}{\begin{fmfgraph*}(5,3)
        \fmfleft{ld,lu}
        \fmfright{rd,ru}
        \fmf{fermion,label=$e^-$}{ld,wl}
        \fmf{fermion,label=$e^+$,label.side=right}{wr,rd}
        \fmf{phantom}{wl,lu}
        \fmf{phantom}{wr,ru}
        \fmf{fermion}{wl,wr}
        \fmf{boson,tension=0,label=$A$,label.side=left}{wr,lu}
        \fmf{boson,tension=0,label=$A$,label.side=right}{wl,ru}
        \fmflabel{$s'$}{lu}
        \fmflabel{$r'$}{ru}
        \fmflabel{$r$}{rd}
        \fmflabel{$s$}{ld}
      \end{fmfgraph*}
      }
    \end{fmffile}.\nb\\
\eqn
Their amplitude reads
\bqn
i{\cal{M}}^{srs'r'}&=&\bar{v}^r(k)\left(g\{\gamma^{\nu},S^{mn}\}\right)u^s(p)\frac{-\frac{i}{2}\eta_{\nu\mu}\left(\eta_{mi}\eta_{nj}-\eta_{mj}\eta_{ni}\right)}{(p+k)^2}
    X^{ij\mu,i_1j_1\mu_1,i_2j_2\mu_2}(p',k')e^{s'*}_{i_1j_1\mu_1}(p')e^{r'*}_{i_2j_2\mu_2}(k')\nb\\
    &+&\bar{v}^r(k)e^{r'*}_{i_2j_2\mu_2}(k')\left(g\{\gamma^{\mu_2},S^{i_2j_2}\}\right)\frac{i(p-p')^{\alpha}\gamma_{\alpha}}{(p-p')^2}
\left(g\{\gamma^{\mu_1},S^{i_1j_1}\}\right)e^{s'*}_{i_1j_1\mu_1}(p')u^s(p)\nb\\
    &+&\bar{v}^r(k)e^{s'*}_{i_1j_1\mu_1}(p')\left(g\{\gamma^{\mu_1},S^{i_1j_1}\}\right)\frac{i(p-k')^{\alpha}\gamma_{\alpha}}{(p-k')^2}
\left(g\{\gamma^{\mu_2},S^{i_2j_2}\}\right)e^{r'*}_{i_2j_2\mu_2}(k')u^s(p).
\eqn
\begin{figure}[h]
\centering
\includegraphics[width=14cm]{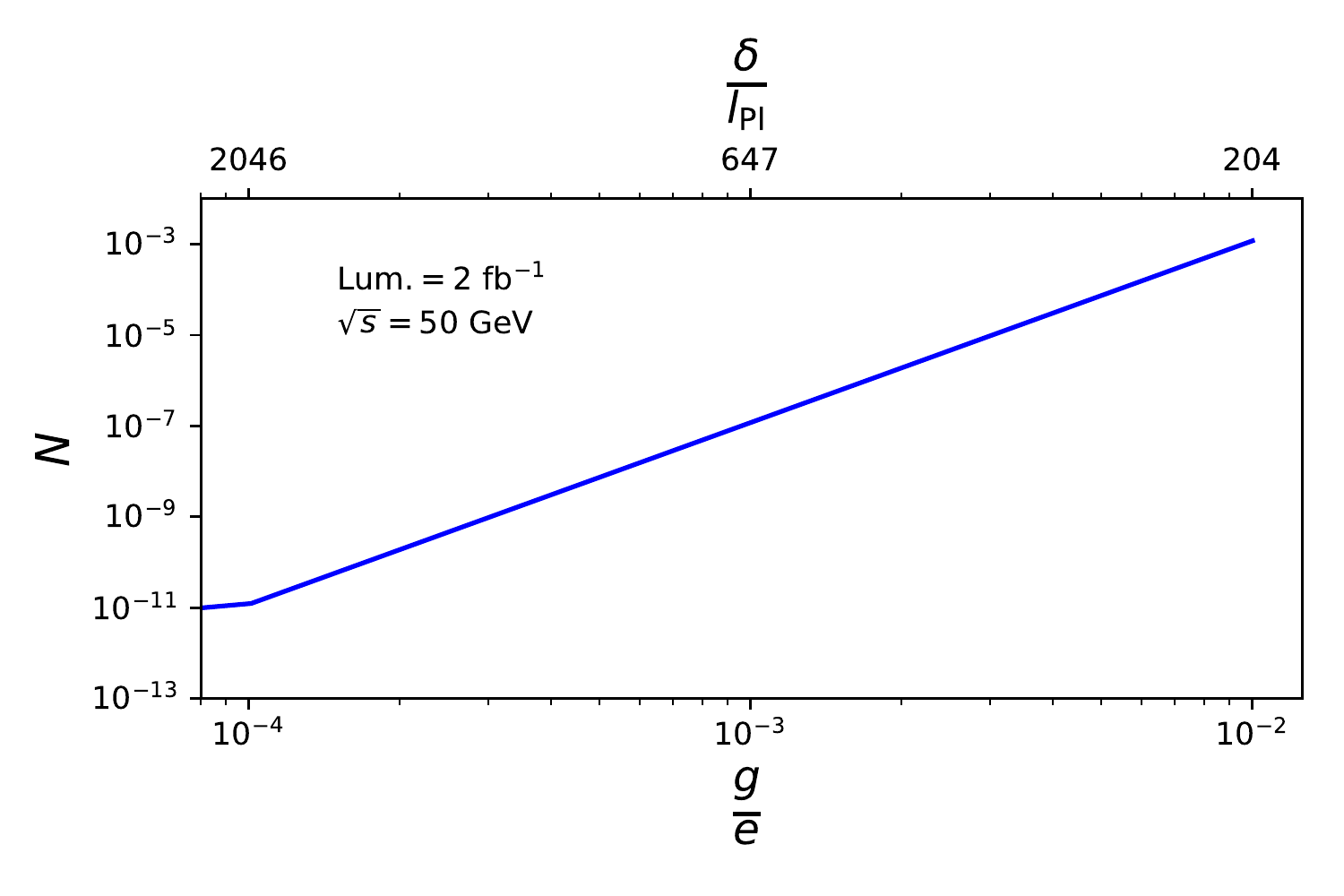}
\caption{Total number of pair annihilation into two gravitons in a particle detector with pseudorapidity coverage of $|\eta|\leq2.44$ after collecting 2 fb$^{-1}$ data at a center of mass energy of 50 GeV.}
\lb{NPairAnnihilation}
\vspace{0.75cm}
\end{figure}
The polarized amplitudes are found to be
\begin{align}
&{\cal{M}}^{1221}=-\frac{1}{2}g^2\left( -2 + \left(2+i \right)\cos\left( \theta \right) -  \cos\left(2 \theta \right)  +i \sin\left( \theta \right) + \tan\left(\frac{\theta}{2} \right)      \right),&~\nb\\
&{\cal{M}}^{2112}=\frac{1}{2}g^2\left( 2 - \left(2-i \right)\cos\left( \theta \right) +  \cos\left(2 \theta \right)  +i \sin\left( \theta \right) - \tan\left(\frac{\theta}{2} \right)      \right),&~\nb\\
&{\cal{M}}^{1212}=\frac{1}{2}g^2\left( 2 + \left(2+i \right)\cos\left( \theta \right) +  \cos\left(2 \theta \right)  +i \sin\left( \theta \right) + \cot\left(\frac{\theta}{2} \right)      \right),\nb\\
&{\cal{M}}^{2121}=\frac{1}{2}g^2\left( 2 + \left(2-i \right)\cos\left( \theta \right) +  \cos\left(2 \theta \right)  -i \sin\left( \theta \right) + \cot\left(\frac{\theta}{2} \right)      \right),\nb\\
&{\cal{M}}^{1211}=\frac{1}{2}g^2\left( i + \cos\left( 2\theta \right)\right),\nb\\
&{\cal{M}}^{1222}=\frac{1}{2}g^2\left( -i + \cos\left( 2\theta \right)\right) ,\nb\\
&{\cal{M}}^{2111}=\frac{1}{2}g^2\left( i + \cos\left( 2\theta \right)\right),\nb\\
&{\cal{M}}^{2122}=\frac{1}{2}g^2\left( -i + \cos\left( 2\theta \right)\right).
\end{align}
The unpolarized amplitude is
\bqn
\frac{1}{4}\sum_{\text{spin}}|{\cal{M}}|^2=
\frac{g^4}{16}\left(\frac{18\left(1+\sin\left(2\theta\right)\right)-5\cos\left(2\theta\right)-4\cos\left(4\theta\right)-\cos\left(6\theta\right)+\sin\left(4\theta\right)}
{\sin^2\left(\theta\right)}\right),
\eqn
which is singular at $\theta=0$ due to neglecting electron mass. The singular point is however out of pseudorapidity coverage of most particle detectors. Therefore, we can leave the singularity out and integrate the rest to get the total number of such events. Assuming that a given detector covers only the pseudorapidities $|\eta| \leq 2.44$, the total number of created graviton pairs after collecting 2 fb$^{-1}$ data at center of mass energy of 50 GeV is drawn in figure \ref{NPairAnnihilation} as a function of the coupling constant of LGT over that of QED, $r$.

\section{Conclusions}
\lb{conclusionSec}
The number of divergences in a renormalizable theory of gravity remains limited, no matter what order of perturbation is desired. This is met only if the coupling constant of the theory does not have a negative dimension. Fortunately, LGT has a dimensionless coupling constant and is expected to have a good high energy behavior which makes it a viable candidate for quantum theory of gravity. In LGT, not only mass of fermions but also their spin gravitate. The mass-generated gravity is expected to become significant only at distance scales as small as the Planck length. Nevertheless, studies suggest that this is not true for the spin-generated gravity that is important at very lower scales. In this paper we have investigated the observable signals of the latter type of interaction between matter and gravity. It has been shown that their Feynman vertex diagram can be converted to the QED vertex if the graviton line is replaced by that of a photon. 
Due to this similarity, there are corrections from LGT to QED dominated processes. Since electron-positron colliders are very popular, only those effects of LGT that can be observed in the data from such accelerators have been studied.  
We have specifically studied $e^-+e^+\rightarrow \mu^-+\mu^+$ and shown that unlike in QED where the differential cross section is symmetric between the backward and forward detector hemispheres, in LGT fermions tend to scatter more into the backward region. Therefore, an asymmetry between the number of events in the two hemispheres can be interpreted as a signal of LGT. Bhabha scattering has been studied as well. 
The two studies suggest that the coupling constant of LGT is at least three orders of magnitude smaller than that of QED. But, it is still far from the Planck scale. 

Gravitational plane waves have been investigated as well. We have shown that these waves are transverse just like electromagnetic counterparts and have derived the corresponding Feynman diagrams. 
Pair annihilation into gravitons is the last studied process. We have shown that the cross section of such processes is much smaller than those of the other two QED dominated ones. Moreover, since current or near future particle detectors are blind to gravitons, the two out-going particles will just disappear without further track.


\section*{Appendix A:  A Conserved Current Made of Fermions}
\renewcommand{\theequation}{A.\arabic{equation}} \setcounter{equation}{0}

Normally any invariance comes with a conserved current which can be found by infinitesimally transforming the action of matter and setting that to zero
\bqn
\lb{Lmatter}
\delta \int \sqrt{-g}d^4x {\cal{L}}_{\text{M}} =0.
\eqn   
For the sake of simplicity we assume that matter is described by a Dirac Lagrangian plus the vacuum energy, instead of using the standard model Lagrangian,
\bqn
\lb{MatterLagrange}
{\cal{L}}_{\text{M}}=\frac{i}{2}\bar{\psi}\gamma^ie_i^{~\mu}D_{\mu}\psi - \frac{i}{2}D_{\mu}\bar{\psi}\gamma^ie_i^{~\mu}\psi +
\text{constant},
\eqn
where 
\bqn
&&D_{\mu}\psi=\partial_{\mu}\psi-gS^{mn}A_{mn\mu} \psi,\nb\\
&&D_{\mu}\bar{\psi}=\partial_{\mu}\bar{\psi}+g\bar{\psi}S^{mn}A_{mn\mu},
\eqn
and $S^{mn}=\frac{1}{4}\left[\gamma^m,\gamma^n\right]$ refers to the generators of the Lorentz group. 
Now we need to see how different objects transform under infinitesimal Lorentz transformations in the tangent spaces
\bqn
&&\delta \psi = S^{mn}\omega_{mn}\psi,\nb\\
&&\delta A_{mn\mu}=D_{\mu}\omega_{mn},\nb\\
&&\delta e_{m\mu}=\omega_{mn}e^n_{~\mu},\nb\\
&&\delta g_{\mu\nu}=0,
\eqn
where $\omega_{mn}$ is an antisymmetric arbitrary parameter. The metric remains unchanged under Lorentz transformations because it purely belongs to space-time. To see the consistency of the equations
\bqn
\delta g_{\mu\nu} &=& \eta^{mn}\left(\delta e_{m\mu}e_{n\nu} +   e_{m\mu}\delta e_{n\nu}\right)\nb\\
&=&e_{m\mu}e_{n\nu} \left( \omega^{mn} +  \omega^{nm}\right),
\eqn
which is zero because $\omega^{mn}$ is antisymmetric. Since the metric compatible Christoffel symbols and the determinant of the metric are made of the metric, they also remain unchanged under such transformations
\bqn
&&\delta \sqrt{-g}=0,\nb\\
&&\delta \Gamma^{\alpha}_{\mu\nu}=0.
\eqn
An important consequence of the latter equation is that the vacuum energy will have no contribution to the conserved source. 
Inserting everything into equation \eqref{Lmatter}
\bqn
\int \sqrt{-g}d^4x \delta\left[
\frac{i}{2}\bar{\psi}\gamma^ie_i^{~\mu}\left(\partial_{\mu}-S^{mn}A_{mn\mu}\right)\psi -
\frac{i}{2}\bar{\psi}\left(\cev{\partial}_{\mu}+S^{mn}A_{mn\mu}\right)\gamma^ie_i^{~\mu}\psi
\right]=0.
\eqn
Also $\frac{\partial {\cal{L}}_{\text{M}}}{\partial \psi}\delta\psi=\delta\bar{\psi}\frac{\partial {\cal{L}}_{\text{M}}}{\partial \bar{\psi}}=0$ are just the Dirac field equations and can be removed. The latter equation therefore can be written as 
\bqn
\lb{LGTSourceDerivation}
&&\int \sqrt{-g}d^4x \left[ i\bar{\psi}\gamma^i\delta e_i^{~\mu}D_{\mu}\psi- i\bar{\psi} e_i^{~\mu}\{\gamma^i,S^{mn}\}\delta A_{mn\mu}\psi\right]=\nb\\
&&\int \sqrt{-g}d^4x \left[ \frac{\delta {\cal{L}}_{\text{M}}}{\delta e_i^{~\mu}} \omega_{ij} e^{j\mu} - i\bar{\psi} e_i^{~\mu}\{\gamma^i,S^{mn}\}D_{\mu}\omega_{mn}\psi\right]=0.
\eqn
It should be noted that the tetrad postulate can be solved such that the tetrad is expressed in terms of the spin connection, at least perturbatively. Therefore, we could in principle write $\delta e_{i\mu}= \frac{\delta e_{i\mu}}{\delta A_{mn\nu}}\delta A_{mn\nu}$. Nevertheless, a better approach is to define $S^{\mu\nu i}$ such that $D_{\nu}S^{\nu\mu i}=\frac{\delta {\cal{L}}_{\text{M}}}{\delta e_{i\mu}}$. 
This is in fact the Lagrange multiplier in equations \eqref{LagrangeMulti} and \eqref{fieldconstraint}. Inserting this back to the equation and using the tetrad postulate $D_{\mu}e_{i\nu}=0$ and consequently $D_{\mu}g=0$
\bqn
\int \sqrt{-g}d^4x D_{\mu}\left(  S^{\mu n m}+ i\bar{\psi}e_i^{~\mu}\{\gamma^i ,S^{mn}\}\psi\right)\omega_{mn}=0.
\eqn
Since $\omega_{mn}$ is an arbitrary antisymmetric parameter, the term multiplied to that has to be zero
\bqn
D_{\mu}\left(  S^{\mu [n m]}+ i\bar{\psi}e_i^{~\mu}\{\gamma^i ,S^{mn}\}\psi\right)=0.
\eqn
Finally the conserved current is
\bqn
J^{\mu mn}=-S^{\mu [m n]}+ i\bar{\psi} e_i^{~\mu}\{\gamma^i,S^{mn}\}\psi.
\eqn

\end{document}